\documentclass{PoS}
\usepackage{amsfonts}
\usepackage{amssymb}
\usepackage{psfig}
\usepackage{mathrsfs}
\PoS{PoS(LAT2005)327}
\newcommand{\beq}{\begin{equation}}
\newcommand{\eeq}{\end{equation}}
\newcommand{\beqn}{\begin{eqnarray}}
\newcommand{\eeqn}{\end{eqnarray}}
\newcommand{\bea}[1]{\beq\begin{array}{#1}}
\newcommand{\eea}{\end{array}\eeq}

\newcommand{\Tr}[1]{\;{1\over #1}\mathop{\rm Tr}}
\newcommand{\tr}{\mathop{\rm Tr}}

\newcommand{\Pexp}{\mbox{P}\!\exp}

\title{Bianchi Identities and Degeneracy of Chromomagnetic Fields in SU(2) Gluodynamics}
\ShortTitle{Bianchi Identities and Degeneracy of Chromomagnetic Fields in SU(2) Gluodynamics}

\author{\speaker{F.V.~Gubarev}, S.M.~Morozov, \\
         ITEP, B.Cheremushkinskaya 25, Moscow, 117218, Russia, \\
         E-mail: \email{gubarev@itep.ru}, \email{smoroz@itep.ru}
}
\FullConference{XXIIIrd International Symposium on Lattice Field Theory\\
                 25-30 July 2005\\
                 Trinity College, Dublin, Ireland}
\abstract{
We investigate the non-Abelian Bianchi identities in pure SU(2) lattice Yang-Mills theory in 
three and four dimensions.
The non-Abelian Stokes theorem proposed recently allows to formulate the Bianchi identities
in terms of local physical fluxes. Then the violation of Bianchi identities  becomes a well
defined concept ultimately related to chromomagnetic fields degeneracy points.
We present numerical evidences that in D=4 the suppression  of the Bianchi identities violation
destroys confinement while the removal of the degeneracy points
drives the theory to the topologically non-trivial sector.}

\begin{document}
\section*{Bianchi Identities and Chromomagnetic Fields Degeneracy Points on the Lattice}

The possibility of Bianchi identities violation had always been considered as probable
cause of confinement (see, e.g., Ref.~\cite{Polyakov} for recent discussion).
However, detailed treatment of the problem in non-Abelian case was missing
until recently~\cite{self-main} and here we briefly review the main points of 
this paper. Formally, the Bianchi identities in three and four dimensions
\beq
\label{BI-continuum}
D_\mu \, \tilde{F}_{\mu\nu} = 0\,,\,\,\,\,
\tilde{F}^a_{\mu\nu} = \frac{1}{2} \varepsilon_{\mu\nu\lambda\rho} F^a_{\lambda\rho}
\,\,\,\,(4D)\,,\qquad
D_i\,B_i = 0\,,\,\,\,\,
B^a_i = 1/2\,\varepsilon_{ijk}\,F^a_{jk}
\,\,\,\,(3D)
\eeq
look similar, but in fact there is a crucial difference: for given non-Abelian field strength
Eq.~(\ref{BI-continuum}) allows to express the gauge potentials as local single-valued functions of
$F_{\mu\nu}$, while in three dimensions it does not constraint $A_\mu$ at all.
The inversion $A(F)$ is possible away from the points chromomagnetic fields degeneracy~\cite{degeneracy},
where $\mathrm{det}\, T = 0$, $T^{ab}_{\mu\nu} = \varepsilon^{abc}\, \tilde{F}^c_{\mu\nu}$.
Note that
\beq
\label{DET-continuum}
\mathrm{det}\, T \propto \mathrm{det}\, K\,,\qquad
K_{\mu\nu} = K_{\nu\mu} =
\frac{1}{3}\,\varepsilon^{abc} \,\tilde{F}^a_{\mu\rho}\, F^b_{\rho\lambda}\, \tilde{F}^c_{\lambda\nu}
\eeq
and each element of $K_{\mu\nu}$ is gauge invariant determinant constructed from chromoelectric $E_i^a$
and chromomagnetic $B_i^a$ fields.
It is amusing that our investigation of the Bianchi identities (\ref{BI-continuum})
and possibility of their violation naturally leads to the same determinants (\ref{DET-continuum}).
Moreover, the degeneracy points turn out to be closely related to the gauge fields topology.

We start from the observation~\cite{Kiskis-1} that the Bianchi identities (\ref{BI-continuum}) are
equivalent to the requirement that the non-Abelian Stokes theorem (NAST) being applied to closed
infinitesimal surface $S_0$ always gives unity on the r.h.s. (surface independence)
\beq
\label{BI-NAST}
\mbox{P}_S \exp\,\frac{i}{2}\int\limits_{S_0, \delta S_0 = 0} \mathscr{F}_{\mu\nu} \, d^2 \sigma^{\mu\nu}
= 1 = e^{\,i\, \vec{\sigma}\vec{n}\cdot 2\pi q}\,,
\qquad \vec{n}^2 = 1\,,
\qquad q\, \in \,Z\,,
\eeq
where for definiteness we used the operator formulation~\cite{NAST-operator} of NAST
and r.h.s. was represented in most general form. The color direction $\vec{n}$ is gauge
variant and is of no concern. However, the possibility to give 
unambiguous gauge invariant meaning to the integer $q$ (``magnetic charge'') is important
since if it is non vanishing then the continuum Bianchi identities are violated somewhere inside $S_0$.
Note that the term ``magnetic charge'' here has nothing to do with its usual meaning, in particular,
its conservation is neither assumed nor implied (see also below).
Since Eq.~(\ref{BI-NAST}) is the definition of the magnetic charge the only way
to make sense of $q$ is to express the non-Abelian Stokes theorem in terms of 
gauge invariant physical fluxes.

The clue to the above problem is to choose the appropriate parametrization~\cite{self-main,Gubarev}
for the Wilson loops
\beqn
\label{wilson}
W(C,t) = \Pexp i\int^{T+t}_t A(\tau)d\tau = \exp\{i\vec{\sigma}\vec{n}(C,t)\cdot\Phi(C)\}\,, \\
\vec{n}^2(C,t) = 1\,, \qquad W(C) = \Tr{2} W(C,t) = \cos\Phi(C)\,, \nonumber
\eeqn
where $C=\{x(t),0\le t\le T,x(0)=x(T)\}$ is some closed contour and it is understood that
$W(C) \ne \pm 1$ generically. The flux magnitude $\Phi(C) \in [0;\pi]$ is gauge invariant,
while the instantaneous flux color direction $\vec{n}(C,t)$ explicitly depends on $t$
and is in the adjoint representation. Note that the flux magnitude is insensitive
to the change of contour orientation while the flux direction changes sign.
Consider now another contour $C'$ which touches
(or intersects) $C$ at point $x(t)=x'(t')$. Evidently, the relative orientation (angle in between)
of $\vec{n}(C,t)$ and $\vec{n}(C',t')$ remains intact under the gauge transformations.
Moreover, the construction could be iterated: for $N$
contours $C_i$, $i=1,...,N$ intersecting at one point the relative orientation
of instantaneous fluxes $\vec{n}^{(i)}$ at that point is gauge invariant. In particular,
the oriented area\footnote{
The total area of unit two-dimensional sphere is taken to be $2\pi$.
} $\Omega_N(\vec{n}^{(1)},...,\vec{n}^{(N)})$ of spherical N-vertex polygon is well defined.

The next point concerns the relation between the physical field strength and
the corresponding infinitesimal Wilson loop.
The basic and well known fact is that the lattice area element $dx^\mu dx^\nu$ is unoriented
$dx^\mu dx^\nu = dx^\nu dx^\mu$ contrary to the usual continuum relation
$\delta\sigma^{\mu\nu} = dx^\mu \wedge dx^\nu = - dx^\nu \wedge dx^\mu$.
Therefore in order to define the lattice counterpart of the field strength
a canonical orientation of all elementary plaquettes should be fixed first.
Moreover, the canonical ordering is well known, the conventional agreement
is to consider $\delta\sigma^{\mu\nu}$ with $\mu < \nu$ only.
{}From now on we always assume that the infinitesimal fluxes are constructed
with canonically oriented plaquettes.

Once the canonical orientation is fixed, the NAST presented and discussed in details
in Ref.~\cite{Gubarev} becomes completely unambiguous. It relates the flux piercing the large
contour $C$ with magnitudes and relative orientations of elementary fluxes on (arbitrary)
surface bounded by $C$. The requirement of surface independence (\ref{BI-NAST}) considered
for elementary lattice 3-cube $c$ becomes then
\beq
\label{bianchi-lattice}
\sum\limits_{p\in \delta c} I(p)\,\Phi(p) +
\sum\limits_{x\in \delta c} \Omega_3(\vec{n}^{(1)}_x,\vec{n}^{(2)}_x,\vec{n}^{(3)}_x) = 2\pi\,q(c)\,,
\eeq
where $\Phi(p)$ is the plaquette flux, $1/2\tr W(p) = \cos\Phi(p)$, $\Omega_3$ was discussed above and accounts for
the difference in color orientations of three fluxes emanating from 3-cube and having one common
point $x\in \delta c$. The factors $I(p) = \pm 1$ are  the usual incidence numbers:
$I(p) = 1$ if vertices of $p$ are followed in the canonical order
and $I(p)=-1$ otherwise.
Eq.~(\ref{bianchi-lattice}) is the definition of the magnetic charge associated with 3-cube
and indeed expresses the essence of non-Abelian Bianchi identities. In particular, the continuum
Bianchi identities (\ref{BI-continuum}) are violated inside $c$ if $q(c)\ne 0$.

It is important that the non-Abelian Bianchi identities (\ref{BI-continuum}) became completely
Abelian-like (\ref{bianchi-lattice}) in our approach. Moreover, it is possible to construct
a specific but unique cell complex in which every term on l.h.s. of Eq.~(\ref{bianchi-lattice})
is associated with particular 2-dimensional cell. Then Eq.~(\ref{bianchi-lattice})
is the usual coboundary operation $d: \mathbb{C}^2\to\mathbb{C}^3$ acting on particular 2-cochains.
One could say that the non-Abelian nature of (\ref{BI-continuum}) had been traded for much
more complicated geometry underlying (\ref{bianchi-lattice}), but which allows purely formal investigation.
In particular, it is clear that the 2- and 3-skeletons of the cell complex are not exhausted by
the original plaquettes and 3-cubes
(see Ref.~\cite{self-main} for details). Then the study of  $d: \mathbb{C}^2\to\mathbb{C}^3$
in its generality leads to the consideration of the ``magnetic charges'' assigned to all the 3-cells
of the complex. It is true that some of these ``new'' 3-cells are trivial and the corresponding
magnetic charge is identically zero. However, there are a non-trivial 3-cells as well which
are distinct from the original lattice 3-cubes. One could show that the corresponding
charges  $\tilde{q}$  signal the zeros of the determinants entering (\ref{DET-continuum}).
Formally the charges $\tilde{q}$ are coming from the $\Omega$ terms
in (\ref{bianchi-lattice}) and indeed depend only upon flux color directions, not their magnitudes.
Note that the union of all magnetic charges $q$, $\tilde q$ is indeed conserved, but the conservation
takes place not in the original hypercubical geometry.

To summarize, the attempt to consider gauge invariant content of the Bianchi identities
(\ref{BI-continuum}) forces us to introduce two types of the ``magnetic charges'' $q$, $\tilde{q}$. First one,
Eq.~(\ref{bianchi-lattice}), is responsible for the Bianchi identities violation and is assigned to
lattice 3-cubes. Second one, being assigned to lattice sites, is directly related to the degeneracy points,
where a particular determinants constructed from $E^a_i$, $B^a_i$ vanish.

\subsection*{Numerical Experiments}

Simple weak coupling analyzes show that in perturbation theory the $q$, $\tilde{q}$ charges
form tightly bounded dipole pairs, but cannot annihilate because at any finite lattice spacing
they by definition are finitely separated. This invalidates the consideration of the corresponding
densities since they both would diverge in the continuum limit.
On the other hand, the definition of $q$ and $\tilde q$ charges is local and gauge invariant. Therefore,
nothing prevents us from modifying the Wilson action to include additional terms which
suppresses the charges $q$, $\tilde q$ 
\beq
\label{action}
S = -\beta\,\sum\limits_p \,\Tr{2}U_p + \gamma \, \sum\limits_c \, |q(c)|
+ \tilde{\gamma} \,\sum\limits_s \,|\tilde{q}(s)|
\eeq
and allows to study their physical relevance. 
The particular limit $\gamma \to \infty$ is of special interest since it corresponds to
the theory with nowhere violated Bianchi identities. As far as the $\tilde{\gamma}$ coupling
is concerned the limit $\tilde{\gamma} \to \infty$ might not correspond to sensible theory.
For instance, the nowhere vanishing $\mathrm{det}\,B$ implies that it is of the same
sign everywhere, which contradicts the perturbative expectations
and probably violates $CP$ symmetry.

\begin{figure}
\centerline{
\psfig{file=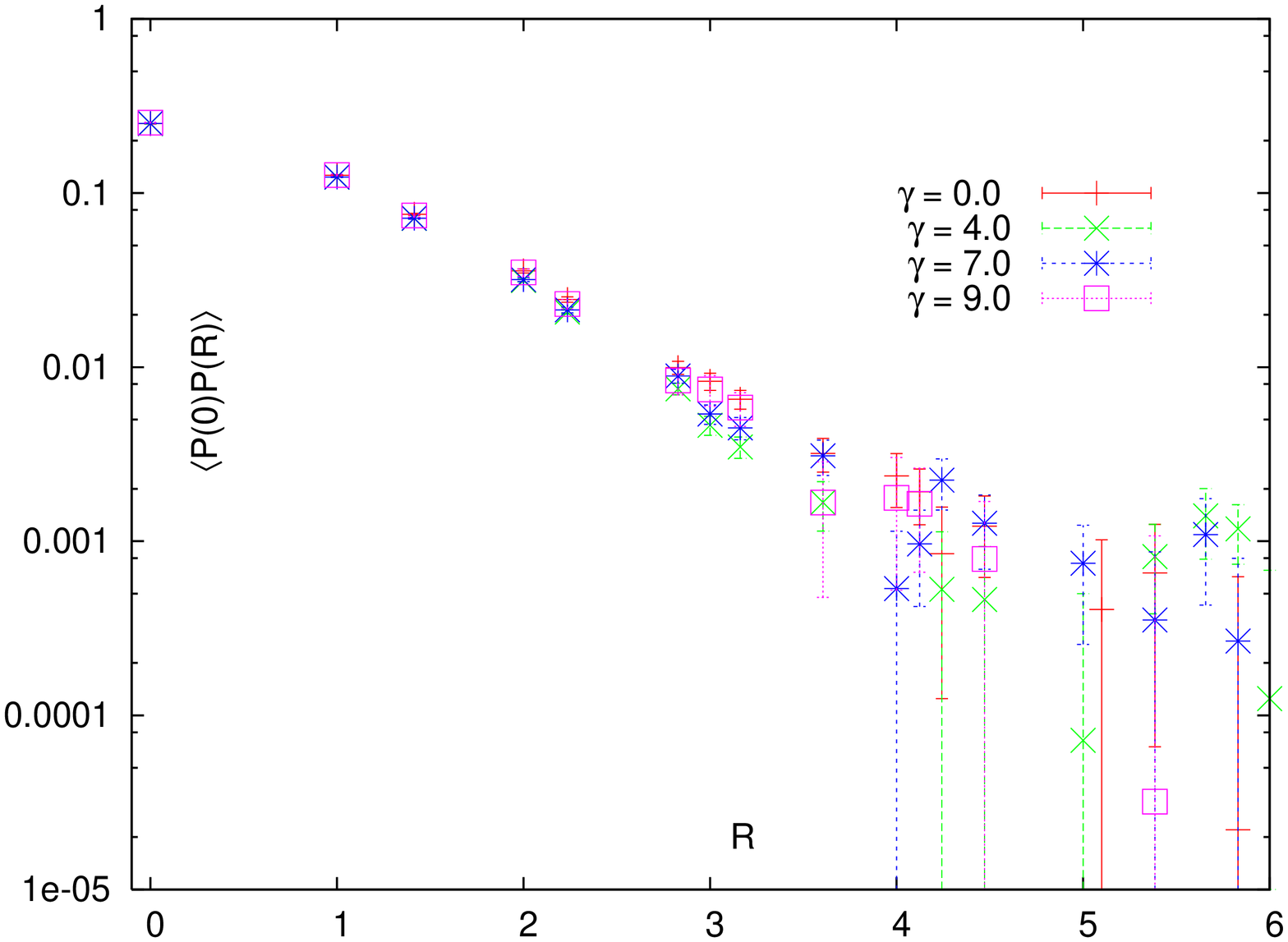,width=0.55\textwidth,silent=,angle=-90}
\psfig{file=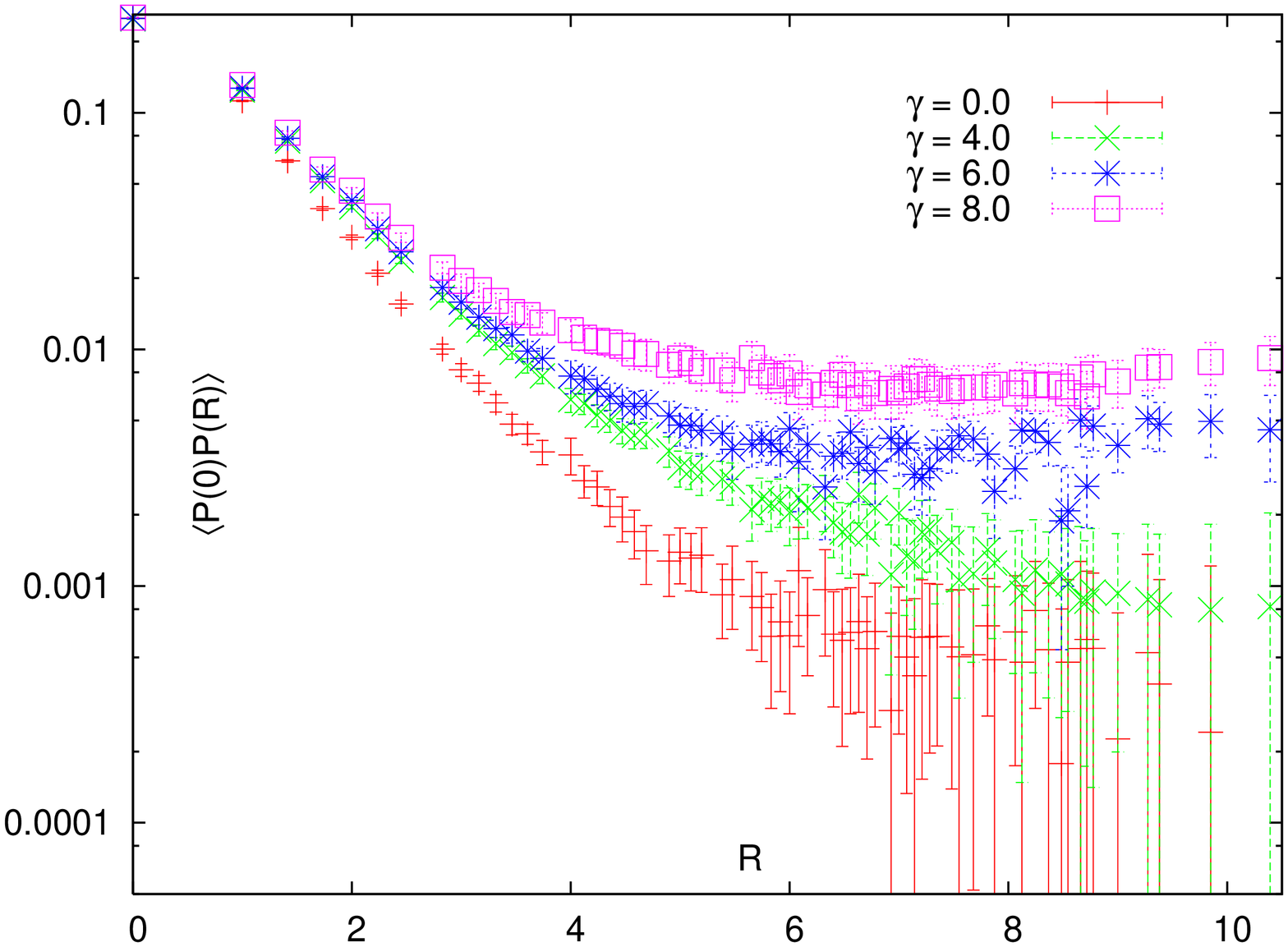,width=0.55\textwidth,silent=,angle=-90}}
\caption{The correlation function of Polyakov lines at $\tilde\gamma = 0$ and
various $\gamma$ in three (left) and four (right) dimensions.}
\label{fig1}
\end{figure}

The model (\ref{action}) was studied in $D=3$, $4$ along the lines $\tilde{\gamma} = 0$,  $\gamma = 0$
at fixed value of the gauge coupling $\beta$.
The simulations were performed on $20^3$ and $12^4$ lattices at $\beta=6.0$ and $\beta=2.4$ correspondingly.
We implemented the Metropolis algorithm which is the only one available at non-zero
$\gamma$, $\tilde{\gamma}$. The procedure turns out to be very time consuming.
Because of this the volumes were not so large and only the limited set of $(\gamma,\tilde\gamma)$
points were considered.
At each $(\gamma,\tilde\gamma)$ values we generated about one hundred statistically independent gauge samples
separated by $\sim 10^3$ Monte Carlo sweeps. The observable of primary importance is
the correlator of the Polyakov lines $\langle P(0)P(R)\rangle$ for which
the standard spatial smearing~\cite{Albanese:1987ds} and hypercubic
blocking~\cite{Hasenfratz:2001hp} for temporal links were used.
In D=4 we also monitored the topological charge $Q$ and
the topological susceptibility $\chi = \langle Q^2 \rangle /V$
measured with overlap Dirac operator~\cite{Neuberger}
(see, e.g. Ref~\cite{Giusti:2002sm} for review and further references).

\begin{figure}
\centerline{
\psfig{file=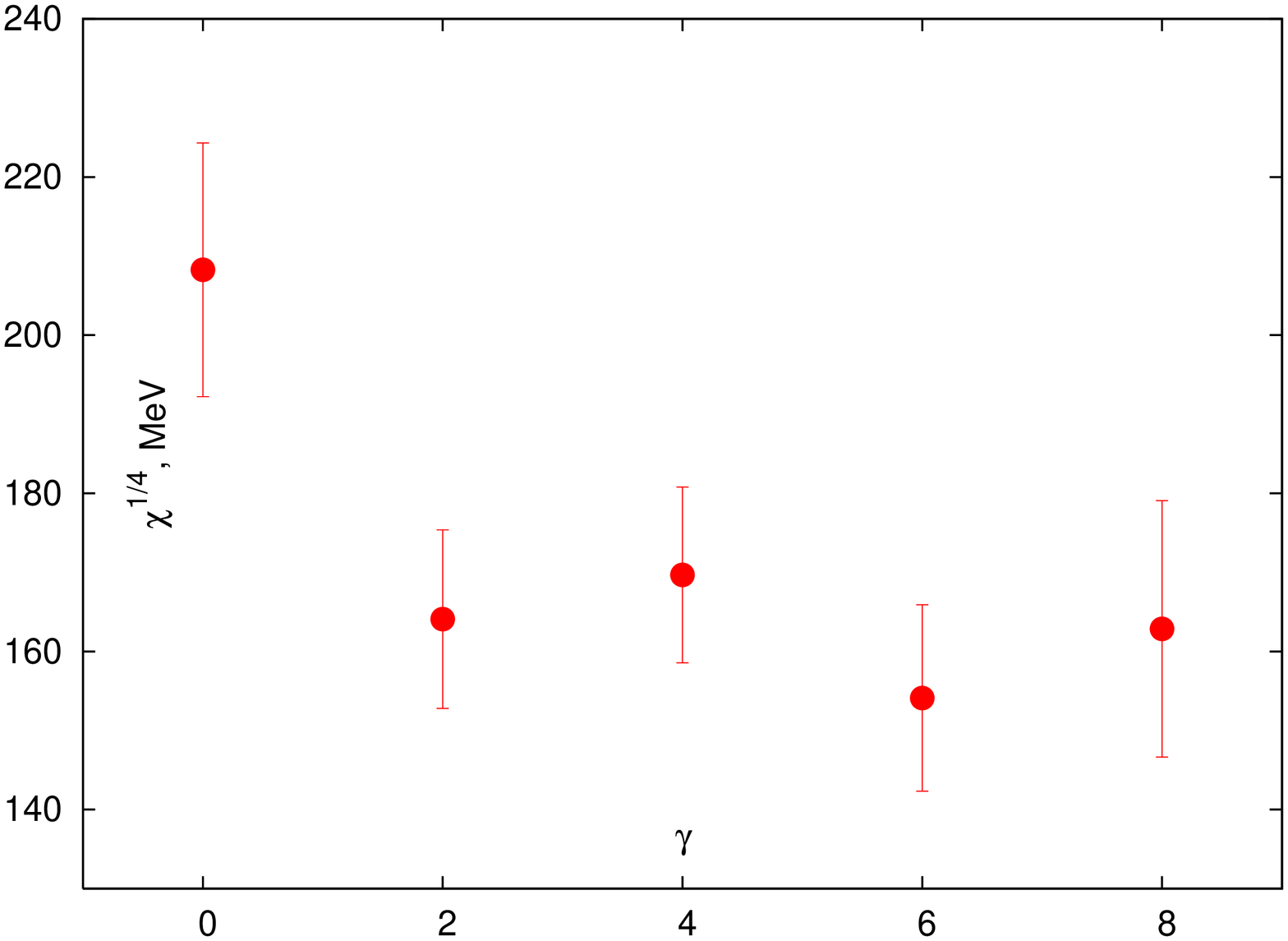,width=0.55\textwidth,silent=,angle=-90}
\psfig{file=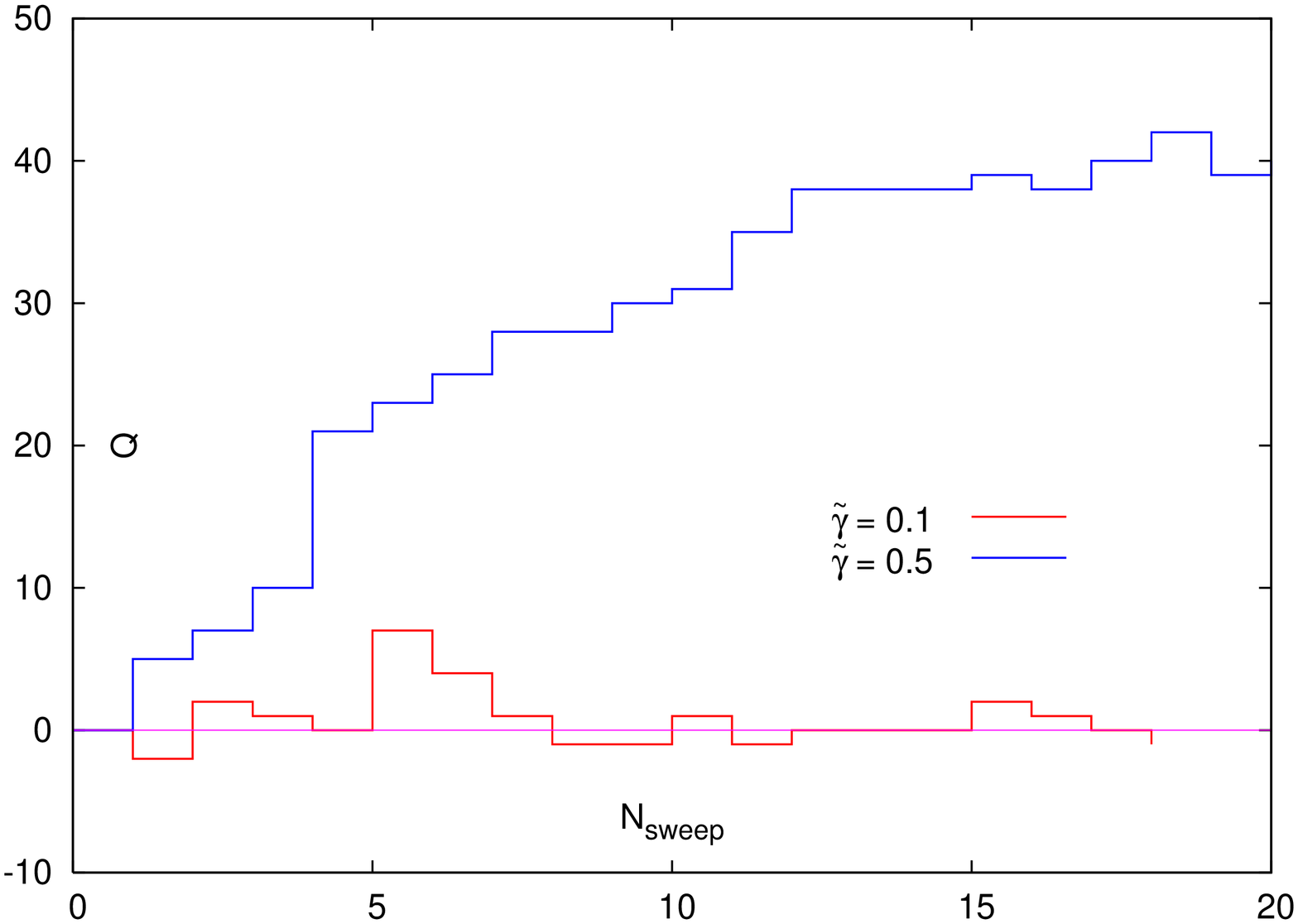,width=0.55\textwidth,silent=,angle=-90}}
\caption{Left: topological susceptibility in D=4  at $\tilde{\gamma} = 0$ and various $\gamma$.
Right: Monte Carlo history of global topological charge at $\gamma = 0$ and two $\tilde\gamma$
values.}
\label{fig2}
\end{figure}

Consider the response of the heavy quark potential on rising $\gamma$ coupling.
In three dimensions (Fig.~\ref{fig1}, left) the correlator $\langle P(0)P(R) \rangle$
shows almost no sign of $\gamma$ coupling dependence, in particular, the asymptotic string
tension at large $\gamma$ is equal to its value in the pure Yang-Mills limit.
However, the situation changes drastically in D=4. It is apparent (Fig.~\ref{fig1}, right)
that the correlation function $\langle P(0)P(R) \rangle$
tends to non-zero positive value at large separations
\beq
\label{PP}
\lim\limits_{R\to\infty} \langle P(0)P(R) \rangle_{\gamma{\scriptscriptstyle\gtrsim} 1} ~=~ const \, > 0\,,
\eeq
when $\gamma$ coupling becomes of order few units. On the other hand,
other measured observables do not show strong dependence on $\gamma$ coupling.
In particular, the topological charge $Q$ stays at zero in average albeit with slightly
narrower distribution. As is clear from Fig.~\ref{fig2} (left) the topological
susceptibility drops at $\gamma \approx 1$ by approximately 25\%
and then stays almost constant, $\lim\limits_{\gamma\to\infty} \, \chi^{1/4}(\gamma) ~=~ 163(8)~\mbox{MeV}$.
Therefore the dynamics of YM fields in D=3 seems to be almost insensitive
to whether or not the Bianchi identities are violated.
Contrary to that, in $D=4$ case the suppression of
the Bianchi identities violation seems to destroys confinement while other measured characteristics
of the theory remain essentially unchanged. For interpretation and more details see Ref.~\cite{self-main}.

As we noted already the suppression of  degeneracy points
might not be physically meaningful. Indeed, we found that at any $\tilde\gamma > 0$
the Polyakov lines correlator becomes an oscillating function of the distance
signaling the reflection positivity violation. In four dimensional case
the natural reason of this is non-zero in average global topological
charge. The right panel on Fig.~\ref{fig2} shows the Monte Carlo history of the topological
charge on $8^4$ lattice at $\beta = 2.30$, $\gamma=0$, $\tilde{\gamma}=0.1, 0.5$
when the starting configuration was thermalized at $\gamma=\tilde{\gamma} = 0$
(the lattice geometry and the $\beta$ coupling were changed for reasons to be explained shortly).
In particular, for $0 < \tilde\gamma \ll 1$ the mean topological charge is shifted only slightly
being of order few units. However, once the $\tilde\gamma$ coupling becomes
comparable with unity $Q$ flows away from zero during Monte Carlo updating towards an extremely
large and positive values with almost constant and very high rate. In fact, it quickly becomes
too large to be technically accessible for us and this was essentially the reason to consider
so small lattices here.  The volume dependence of $\langle Q \rangle$ could be inferred
by noting that the last term in (\ref{action}) responsible for the rapid increase of the topological
charge is the bulk quantity. Therefore $\langle Q \rangle$ is expected to be proportional
to the volume at fixed $\tilde\gamma$ although we had not investigated this numerically.

To conclude we note that the numerical results confirming the connection between
confinement and violation of Bianchi identities should be improved in order to be convincing.
In particular, we hope to study elsewhere the larger volumes and $\beta$ coupling.
Nevertheless, our data clearly indicates the relation between the degeneracy points and
the gauge fields topology, which is worth to be further investigated.

\end{document}